# Are Stock Markets Integrated? Evidence from a Partially Segmented ICAPM with Asymmetric Effects.

**AROURI Mohamed El Hedi**[*]

*LEO & EconomiX, University of Paris X,
200, avenue de la République Nanterre 92001, France.*
mohamed-el-hedi.arouri@u-paris10.fr

**Abstract**

In this paper, we test a partially segmented ICAPM for two developed markets, two emerging markets and World market, using an asymmetric extension of the multivariate GARCH process of De Santis and Gerard (1997,1998). We find that this asymmetric process provides a significantly better fit of the data than a standard symmetric process. The evidence obtained from the whole period and sub-periods analysis supports the financial integration hypothesis and suggests that domestic risk is not a priced factor.

**Keywords:** International Asset Pricing, Financial Integration, Emerging Markets, Multivariate GARCH.
**(JEL Classification** : F36; C32; G15)

---

[*] We would like to think Georges Prat and the referees for helpful comments and suggestions which have greatly improved this study.

# 1 INTRODUCTION

Determining the extent to which a national market is integrated in the world stock market is an empirical question which has decisive impact on a number of issues affecting problems that are addressed by financial market theory. If capital markets are fully integrated, investors face common and country-specific risks, but price only common risk factors because country-specific risk is fully diversified. In this case, the same asset pricing relationships apply in all countries and expected returns should solely be determined by global risk factors. In contrast, when capital markets are segmented the asset pricing relationship varies across countries and expected returns would be determined by domestic risk factors. When capital markets are partially segmented, investors face both common and country-specific risks and price them both. In this case, expected returns should be determined by a combination of local and global risk sources. Thus, expected gains from world portfolio diversification and criteria for capital budgeting decisions will be quite different under local, global and mixed pricing.

Empirical papers investigating stock market integration have been mainly limited to developed markets. These papers include, among others, Dumas and Solnik (1995), De Santis and Gerard (1997,1998), Hardouvelis et al. (2002), Aggarwal et al. (2003) and Gerard et al. (2003). The findings of these studies support the financial integration hypothesis of developed equity markets. Recently, some papers have tented to focus on emerging markets, in particular Asian equity markets, partly as a result of their high retes of economic growth and the 1997 Asian crisis. The results of these studies are heterogeneous.

Masih and Masih (1997) show using cointegration methods that the Asian Newly Industrializing Countries of Honk Kong, Singapore, Taiwan and South Korea share long run relationship with developed markets (U.S., Japan, U.K. and Germany). Masih and Masih (1999) apply recent econometric methods including vector error-correction and level VAR models and find similar results. More recently, Masih and Masih (2001) study the dynamic causal linkages amongst international stock markets. They find significant interdependencies between the estabilshed OECD and the emerging Asian Markets. In particular, their results show the leadership of the US and the UK markets both in the short and long term, despite the global financial crash of October 1987. Lim et al. (2003) examine the linkages between stock markets in the Asian region over the period 1988-2002 using non-parametric cointegration techniques and find that there is a common force which brings these markets together in the long run.

In contrast, Roca and Selvanathan (2001) show using different recent econometric techniques that there is no short-term and long-term linkages among the stock markets of Australia, Hong Kong, Singapore and Taiwan. Phylaktis and Ravazzolo (2000) study the potential linkages between stock prices and exchange rate dynamics for a group of Pacific-Basin capital markets and show lack of comovment during the eighties for the free stock markets of Singapore and Hong Kong.

On the other hand, recent works stress the fact that the level to which markets are integrated or segmented is not fixed, but changes gradually over time. Bekaert and Harvey (1995) estimate the degree of integration for developed and emerging markets and show that indeed stock markets become either more or less integrated over time. Liu and pan (1997) find that the US market is more influential than the Japanese market in transmiting returns and volatilities to the Asian markets and that the observed spillover effects are unstable over time and increase substantially after the October 1987 stock mrket crash. Bilson et al. (2000) show that the regional integration among stock markets in Soutk Korea, Taiwan, Thailand, the Philippines and Malaysia is faster than their integration within the international market. Hooever, Barari (2003) compare the status of regional vis-a-vis global integration of six Latin American equity markets over the period 1988-2001 using time-varying integration score. The empirical evidence shows that integration is time varying and suggests increased global relative to regional stock market integration in recent years. Ratanapakon and Sharma (2002) study the short and long-term relationships in five regional stock indices for the pre-Asian crisis and for the crisis period. They find that the degree of linkage incresed during and after the crisis period. However more recently, Phylaktis and Ravazzolo (2004) apply multivariate cointegration methods to investigate stock market interactions amongst a group of Pacific-Basin countries and the industrialized countries of Japan and US over the period 1980-1998. They show that although the linkages have increased in recent years, there is room for long-term gains by investing in Pacific-Asian markets. In particular, their results show that the Asian crisis did not have a substantial effect on the degree of linkages of these markets.

In the current paper, we estimate a partially segmented international capital asset pricing model (ICAPM) where both local and global sources of risk are priced. The main purpose of the paper is to examine the potential integration of two Pacific-Basin countries (Hong Kong and Singapore), U.K. and U.S. with the world market. The two emerging Asian markets included in the study have enjoyed remarkably rapid economic growth in the past decades and are gaining increased influence in the world capital markets. Therefore, the integration of these markets with developed markets deserves closer attention. This issue is checked for the period 19970-2004, and based on the October 1987 stock market crash, using the pre-October 1987 period, and then the post-October 1987 period.

This study is primarily motivated by several reasons. Firstly, most studies that examine the interdependence between international stock markets use cointegration methods and then test indirectly the stock market integration hypothesis. Moreover, a weakness of cointegration methods is that a focus on comparative statics does not take into account the time variation in equity risk premia (see for instance, Harvey (1991) and Bekaert and Harvey (1995)), which may yield confusing and partial results. To accommodate this feature of the data, we estimate a dynamic version of the model, in which both the prices and quantities of risk vary over time. Secondly, since a number of studies have documented that international equity market integration changes over time, the inclusion of a longer sample period permits us to investigate the impact of changes in world stock markts on the degree of integration. Thirdly, empirical papers investigating directly stock market integration have been mainly limited to developed markets and only some papers have tented to focus on emerging markets. As shown obove, the results of these studies are heterogeneous. In this paper, the dynamic ICAPM is estimated using a multivariate GARCH process simultaneously for 5 markets: the world market, 2 developed markets and 2 emerging markets. Finally, if, as is argued in univariate and bivariate cases by Glosten et al. (1993) and Kroner and Ng (1998), the conditional variances and covariances are higher during stock market downturns, the econometric specification should allow for asymmetric effects in variances and covariances. In the current paper, we develop an asymmetric extension of the multivariate GARCH-in-Mean process of De Santis and Gerard (1997,1998). This approach, with sign and size asymmetric effects, allows to the prices of domestic and world market risks, betas and correlations to vary asymmetrically through time.

The rest of the paper is organised as follows. Section 2 presents the model and introduces the econometric methodology. Section 3 describes the data. Section 4 reports the empirical results. Concluding remarks are in section 5.

## 2 THE MODEL AND EMPIRICAL METHODOLOGY

The Capital Asset Pricing Model (CAPM) predicts that the expected excess return on an asset is proportional to its nondiversifiable risk measured by its covariance with the market portfolio. Under the hypothesises of stock market integration and purchasing power parity, an international conditional version of the CAPM can be written as:

$$E(\tilde{R}_{it} / \Omega_{t-1}) - R_{ft} = \delta_{t-1} Cov(\tilde{R}_{it}, \tilde{R}_{Wt} / \Omega_{t-1}), \quad \forall i \quad (1)$$

where $\tilde{R}_{it}$ is the return on asset i between time (t-1) and t, $R_{ft}$ is the return on a risk-free asset and $\tilde{R}_{Wt}$ is the return on the market portfolio. $\delta_{t-1}$ is the price of world market risk and is equal to the world aggregate risk aversion coefficient, see Merton (1980) and Adler and Dumas (1983). All expectations are taken with respect to the set of information variables $\Omega_{t-1}$.

However, many recent studies show that expected returns in most markets are influenced by both global and local risk factors, *i.e.* most markets are neither fully integrated nor completely segmented, see among others, Bekaert and Harvey (1995), Karolyi and Stulz (2002), Carrieri et al. (2002), Gerard et al. (2003) and Barr and Priestley (2004). In this partially segmented framework, expected returns should be determined by two risk factors: global market risk and residual domestic risk.

$$E(\tilde{R}_{it} / \Omega_{t-1}) - R_{ft} = \delta_{t-1} Cov(\tilde{R}_{it}, \tilde{R}_{Wt} / \Omega_{t-1}) + \delta_{di,t-1} Var(\tilde{\eta}_{it} / \Omega_{t-1}), \quad \forall i \quad (2)$$

where $\delta_{di,t-1}$ is the price of domestic risk and $\tilde{\eta}_{it}$ is the market model residual. Thus, ($Var(\tilde{\eta}_{it} / \Omega_{t-1})$) captures the domestic market nondiversifiable risk uncorrelated to world risk;

$$Var(\tilde{\eta}_{it} / \Omega_{t-1}) = Var(\tilde{R}_{it} / \Omega_{t-1}) - Cov(\tilde{R}_{it}, \tilde{R}_{Wt} / \Omega_{t-1})^2 / Var(\tilde{R}_{Wt} / \Omega_{t-1}). \quad (3)$$

Next, consider the econometric methodoloy. Equation (2) has to hold for every asset including the market portfolio. A benchmark system of equations can be used to test the partially integrated conditional ICAPM. For an economy with $N$ risky assets, the following system of pricing restrictions has to be satisfied at each point in time:

$$\tilde{R}_t - R_{ft}\tau = \delta_{t-1}h_{Nt} + \delta_{d,t-1} * q_t + \tilde{\varepsilon}_t \qquad \tilde{\varepsilon}_t / \Omega_{t-1} \sim N(0, H_t) \qquad (4)$$

where $q_t = D(H_t) - (h_{Nt} * h_{Nt})/h_{NNt}$, and $\tilde{R}_t$ denotes the $(N \times 1)$ vector that includes $(N-1)$ risky assets and the market portfolio, $\tau$ an N-dimensional vector of ones. $H_t$ is the $(N \times N)$ conditional covariance matrix of asset returns, $h_{Nt}$ is the $N^{th}$ column of $H_t$ composed of the conditional covariance of each asset with the market portfolio and $h_{NNt}$ the conditional variance of the world market portfolio. $\delta_{d,t-1}$ is the $(N \times 1)$ vector of time-varying prices of domestic risk, $q_t$ is the $(N \times 1)$ vector on nondiversifiable local risk, $D(H_t)$ the diagonal components in $H_t$ and $(*)$ denotes the Hadamard matrix product.

The dynamics of conditional moments are left unspecified by the model. However, it has been shown that securities exhibit volatility clustering and leptokurtosis. Such characteristics are taken into account by ARCH specification. Moreover, if, as is argued in univariate and bivariate cases by Glosten et al. (1993) and Kroner and Ng (1998), the conditional variances and covariances are higher during stock market downturns, the econometric specification should allow for asymmetric effects in variances and covariances. To accommodate this feature of the data, we develop an asymmetric extension of the multivariate GARCH process proposed by De Santis and Gerard (1997). Formally, $H_t$ can be written as follows:

$$H_t = C'C + aa' * \varepsilon_{t-1}\varepsilon'_{t-1} + bb' * H_{t-1} + ss' * \xi_{t-1}\xi'_{t-1} + zz' * \eta_{t-1}\eta'_{t-1} \qquad (5)$$

where $\xi_{it} = \varepsilon_{it}I_{\xi_{it}}$ where $I_{\xi_{it}} = 1$ if $\varepsilon_{it} < 0$ otherwise $I_{\xi_{it}} = 0$,

$\eta_{it} = \varepsilon_{it}I_{\eta_{it}}$ where $I_{\eta_{it}} = 1$ if $|\varepsilon_{it}| > \sqrt{h_{iit}}$ otherwise 0,

$C$ is a $(N \times N)$ lower triangular matrix, $h_{iit}$ is the conditional variance of asset i and *a*, *b*, *s* and *z* are $(N \times 1)$ vectors of unknown parameters.

This parameterisation implies that the variances in $H_t$ depend asymmetrically only on past squared residuals and an autoregressive component, while the covariances depend asymmetrically upon past cross-products of residuals and an autoregressive component. In particular, it guarantees that the conditional variance matrix is definite and positive. We find the symmetric GARCH process of De Santis and Gerard (1997) when $s = z = 0$.

Next, turn to the price of risk. The evidence in Harvey (1991) and De Santis and Gerard (1997) suggests that the price of risk is time varying. Furthermore, Merton (1980) and Adler and Dumas (1983) show the price of world market risk to be equal to the world aggregate risk aversion coefficient. Since most investors are risk averse, the price of risk must be positive. In this paper, we follow De Santis and Gerard (1997), De Santis et al. (2003) and Gerard et al. (2003) and model the dynamics of the risk prices as a positive function of information variables: $\delta_{t-1} = \exp(\kappa'_W Z_{t-1})$ and $\delta_{di,t-1} = \exp(\kappa'_i Z^i_{t-1})$, where $Z$ and $Z^i$ are respectively a set of global and local information variables included in $\Omega_{t-1}$ and $\kappa$ is a set of weights that the investor uses to evaluate the conditionally expected returns. Finally, note that the variables we use to condition the prices of domestic risks are correlated with the degree of openness and development of the local stock markets, then the model allows implicitly the degree of integration to change over time.

Equations (4) and (5) constitute our benchmark model. Under the assumption of conditional normality, the log-likelihood function can be written as follows:

$$\ln L(\theta) = -\frac{TN}{2}\ln(2\pi) - \frac{1}{2}\sum_{t=1}^{T}\ln|H_t(\theta)| - \frac{1}{2}\sum_{t=1}^{T}\varepsilon'_t(\theta)H_t^{-1}(\theta)\varepsilon_t(\theta) \qquad (6)$$

where $\theta$ is the vector of unknown parameters. To avoid incorrect inference due to the misspecification of the conditional density of asset returns the quasi-maximum likelihood (QML) approach of Bollerslev and Wooldridge (1992) is used. Simplex algorithm is used to initialize the process, then the estimation is performed using BHHH algorithm.

# 3 DATA AND PRELIMINARY ANALYSIS

This section serves two purposes. First, we introduce the data we use in our empirical investigation. Second, we show that the data contains features that can be captured with a GARCH model. The dataset includes two distinct groups of data: the returns series and the global and domestic information variables used to condition the estimation.

We use monthly returns on stock indexes for four countries plus a value weighted world market index over the period February 1970– December 2003. Given the aim of the paper, we select two large markets (the United States and the United Kingdom) and two small markets (Hong Kong and Singapore). All the indices are obtained from Morgan Stanley Capital International (MSCI) and include both capital gains and dividend yields. Returns are computed in excess of the 30-day Eurodollar deposit rate obtained from DataStream and expressed in American dollar. Descriptive statistics for the excess returns are reported in Table 1.

Table 1 reveals a number of interesting facts. The Bera-Jarque test statistic strongly rejects the hypothesis of normally distributed returns, which supports our decision to use QML to estimate and test the model. The values of the unconditional correlations are relatively low. The lack of autocorrelation in the return series reveals that we do not need to include an AR correction in the mean equations.

For the squared returns, autocorrelation is detected at short lags, which suggests that GARCH parameterisation for the second moments might be appropriate. Panel E of table I contains the cross-correlations of squared returns between the world and the other countries at different leads and lags. With few exceptions, only the contemporaneous correlations are statistically significant. This evidence suggests that, at least with our monthly data, the croos-market dependence in volatility is not strong and that the diagonal GARCH parameterisation for the second moments is not too restrictive.

Finally, note that empirical research has found support for a time-varying price of risk, see for examples, Harvey (1991), Bekaert and Harvey (1995) and Dumas and Solnik (1995). The price of risk is often modelled as a function of a certain number of instruments, which are designed to capture expectation about business cycle fluctuations. The logic which justifies the use of these instruments is that investors become more risk averse during economic troughs while the market price of risk decrease during expansionary phases of the business cycle. However, the CAPM is a partial equilibrium model and it does not specify state variables that can explain the observed dynamics of the prices of risk. Previous studies used as conditioning information set variables that are connected with the evolution of financial markets. These conditioning instruments are intended to convey the information available to investors.

In order to preserve the comparability between this study and others studies, the choice of global and local information variables is mainly drawn from previous empirical literature in international asset pricing, see Harvey (1991), Ferson and Harvey (1993), Bekaert and Harvey (1995) and De Santis and Gerard (1997,1998). The set of global information includes a constant, the MSCI world dividend price ratio in excess of the 30-day Eurodollar deposit rate (WDY), the change in the US term premium spread measured by the yield on the ten-year US Treasury note in excess of the one-month T-Bill rate (DUSTP), the US default premium measured by the difference between Moody's Baa-rated and Aaa-rated corporate bonds (USDP) and the change on the one month Euro$ deposit rate (DWIR). The set of local information includes a constant, the local dividend price ratio in excess of the local short-term interest rate (LDY), the change in the local short-term interest rate (DLIR) and and the change in industrial production (DIP). Information variables are from MSCI, the International Financial Statistics (IFS) and DataStream and are used with one-month lag relative to the excess returns. Summary statistics for the conditioning information variables, not reported here in order to preserve space but available on request, show that the correlations among the information variables are low. Hence, our proxy of the information set contains nonredundant variables.

# 4 EMPIRICAL RESULTS

We first estimate, over the full period, the model with the symmetric GARCH process of De Santis and Gerard (1997) and then with the asymmetric GARCH process discussed earlier in the paper. Panel A of Table 2 reports the results of a likelihood ratio test of the symmetric versus the asymmetric process. The test rejects the symmetric specification in favor of the asymmetric one. Similar results are given by the Akaike and Schwarz criterions presented in Panel B. Residual statistics reported in Panel C show that average mean residual is closer to zero using the asymmetric specification. To sum up, our findings show that the partially integrated ICAPM with asymmetric GARCH process fits the data better than the symmetric process of De Santis and Gerard (1997).

Table 3 contains parameter estimates and a number of diagnostic tests for the partially segmented conditional ICAPM with asymmetric GARCH process estimated over the full period 1973-2003.

The ARCH coefficients and GARCH coefficients reported in panel B are significant for all assets. This is in line with previous results in the literature. The coefficients *a* are relatively small in size, which indicates that conditional volatility does not change very rapidly. However, the coefficients *b* are large, indicating gradual fluctuations over time. One of the advantages of our approach is to authorize for asymmetric variance and covariance effects. The significant coefficients in the vector *s* imply that the conditional variance is higher after negative shocks for the United States, Singapore and Hong Kong. The significant coefficients in *s* are all positive, which implies that conditional covariances between these countries increase after common negative shocks. In the same way, the significant coefficients in vector *z* indicate that the conditional variance is higher after shocks large in absolute value for the US, UK and world market. The significant coefficients in *z* have the same sign (negative). This result shows that conditional covariances between these stock markets increase after large common negative or positive shocks.

Panel A of Table 3 shows the mean equation parameter estimates and Panel C reports some specification tests. For the world price of risk, the constant and the coefficients of the world dividend price ratio, the term premium and the default premium are significant. The average price of market risk is equal to 3.54 and is highly significant, which is consistent with the findings by earlier studies. On the other hand, the conditional version of the model implies that investors update their strategy using the new available information. Thus, there is no reason to believe that the equilibrium price of risk will stay constant. The robust Wald test for the significance of the time-varying parameters in the price of world market risk rejects the null hypothesis at any standard level. Figure 1 plots the estimated price of world market risk. Risk averse investors should demand higher expected returns at times of high expected risk in the economy. Thus, at times of uncertainty the price of risk should be higher than at times of calm. This seems to be confirmed in the figure 1. In fact, the spikes in the conditional price of risk in figure 1 are associated with the oil crisis (1973-1974), the monetary experiment (1979-1982), the October 1987 crash, the Gulf war (1990), the Asian financial crisis (1997-1998) and the terrorist attacks on US (2001).

As in earlier studies, the point estimates are very noisy. Since we are especially interested in the trend in the series, the Hodrick and Prescott (1986) filter (HP) is used to separate the short-term components from the long-term component. A simple visual inspection of the chart shows that the price of market risk reaches its highest values in the Seventies and the early Eighties. Between 1994 and 2000, it becomes much lower. Finally, the price of world market risk increases significantly in the last years of our sample.

Concerning the prices of domestic residual risk, the results show that none of the estimated coefficients of local information variables are significant. The sample means of the prices of domestic risk are 0.68 for the US, 0.95 for Singapore, 1.12 for the UK and 1.63 for Hong Kong. As expected, the prices of domestic risk are all insignificant. The robust Wald tests confirm these results and suggest that domestic risk is not a priced factor, *i.e.* over the sample period the market considered were fully integrated. In fact, the null hypothesis that the domestic risk price coefficients are jointly equal to zero cannot be rejected at any standard level. This result is confirmed by the single country tests. To sum up, no evidence of financial segmentation is detected over the full period 1973-2003.

Next, we consider a number of robustness tests. To address this issue, we estimate an augmented version of the model that includes, in addition to market and domestic risk, a country specific constant and the local instrumental variables $Z^i$:

$$E(\tilde{R}_{it}/\Omega_{t-1}) - R_{ft} = \alpha_i + \delta_{t-1} Cov(\tilde{R}_{it}, \tilde{R}_{Wt}/\Omega_{t-1}) + \delta_{di,t-1} Var(\tilde{\eta}_{it}/\Omega_{t-1}) + \phi_i' Z_{t-1}^i, \quad \forall i \qquad (7)$$

The inclusion of the country-specific constants can be interpreted as a measure of mild segmentation or as an average measure of other factors that cannot be captured by the model like differential tax treatment. The inclusion of local instrumental variables can be interpreted as a way to test whether any predictability is left in the local information variables after they have been used to model the dynamics of the domestic risk prices. The test results are reported in Table 4. The Wald test indicates that the country intercepts are not jointly different from zero. On the other hand, the null hypothesis that the local information variable coefficients are jointly equal to zero cannot be rejected at any standard level.

Taken together, our results support the financial integration hypothesis and suggest that domestic risk is not a priced factor. These results are consistent with the findings of De Santis and Gerard (1997,1998) and Gerard et al. (2003). However, such conclusion seems to be a strong one. It simply implies that the analysed markets are fully integrated along the entire time period covered in the this study, while many recent studies stress the fact that the level to which stock markets are integrated or segmented is not fixed, but changes gradually over time, see, for examples, Bekaert and Harvey (1995), Liu and pan (1997), Bilson et al. (2000), Ratanapakon and Sharma (2002), Aggarwal et al. (2003), Hunter (2004) and Phylaktis and Ravazzolo (2004). Furthermore, Figure

2, which plots for each market the conditional correlation with the world market, shows that correlations have significantly increased during the recent years at least for the UK and Hong Kong. The increased time-varying correlations explain why the whole sample unconditional correlations reported in Table 1 are quite low and suggest higher degrees of stock market integration. Therefore, it is interesting to investigate whether the inferences have changed over time and whether the results are sensitive to the choice of the sample period.

In the rest of the paper, we explore changes in patterns of dynamic integration among national stock market indices following the October 1987 crash. In fact, as shown in Table 1, the worst month for all stock markets, including the world market, is related to the market crash of October 1987. This is clearly a symptom of international contagion. Many empirical works argued that the interdependencies between international stock markets have increased after the October 1987 crash, see, among others, Jeon and Von Furstenberg (1990), Lau and McLnish (1993), Arshanapalli and Doukas (1993) and Liu and Pan (1997). The increased interdependencies may reflect higher levels of integration and therefore, it is reasonable to question whether the degrees of stock market integration are unduly influenced by the October 1987 crash. So, dividing the full sample period into two sub-periods, up to October 1987 and since October 1987, seems appropriate for examinng the evolution of stock market integration.

On the other hand, in the later half of the 1980s and early years of the 1990s, most of governments gradually liberalized their stock markets. In theory, the liberalization should bring about more integrated local markets with global stock markets. However, market liberalization is neither a necessary nor a sufficient condition for international stock market integration. In fact, other factors may exert an effect, such as information avaibility, accounting standards, liquidity, political and currency risks. On the other hand, there can also be a situation in which foreign investors use alternative instruments, for instance country funds, to enter capital markets with foreign restrictions, see Bekaert (1995), Bekaert and Harvey (2000) and Phylaktis and Ravazzolo (2004). Therefore, It is also natural to question whether the levels of integration have changed as national stock markets have become more liberalized. The second sub-period defined above can also be considered as the post liberalization period. One may expect higher integration between national stock markets in this sub-period.

We re-estimate the model over the sub-periods Febraury 1970-September 1987 and October 1987-December 2003. Since we are especially interested in the significance of the prices of local market risk, robust Wald tests are used to evaluate joint hypotheses on these prices of risk. The results of that exercise are summarized in Table 5. In order to preserve space, mean and variance equation parameter estimates are not reported here but are available on request.

For the sub-period 1970-1987, except for the constant term for Hong Kong, the results show that none of the estimated coefficients of local prices of risk are significant. The sample means of the prices of domestic risk are 0.92, 1.35 and 1.56 respectively for the US, UK and Singapore and are all insignificant. However, for Hong Kong the price of local risk is equal to 2.08 and is siginificant at 5%. These results are confirmed by the robust Wald tests that suggest that domestic risk is not a priced factor for the US, UK and Singapore stock markets and that the null hypothesis that the domestic risk price for Hong Kong is equal to zero is rejected at 5%. However, the Wald test shows that the domestic risk price for Hong Kong is not time varying. Finally, the Wald test can not reject the null hypothesis that the four domestic risk pricees are jointly equal to zero.

To sum up, over the sub-period 1970-1987 we find that the stock markets of the US, UK and Singapore are fully integrated in the world capital market, while we find weak support for the hypothesis that the Hong Kong capital market is partially integrated in the world market.

Next, we consider the second sub-period 1987-2003. For this sub-period, our findings are similar to those obtained for the whole period. In particular, none of the estimated coefficients of local prices of risk are significant. The sample means of the prices of domestic risk are 0.48 for the US, 0.73 for Singapore, 0.89 for the UK and 1.12 for Hong Kong. These prices of local risk are all insignificant. The robust Wald tests confirm these results and suggest that local market risk is not a priced factor. In short, no evidence of segmentation is found over this sub-period.

One can postulate several reasons toward explaining this increased stock market integration. Firstly, the strong economic links among the countries analyzed in this study, especially trade and investment that have indirectly linked their equity markets. In fact, economic linkages between countries imply a comovement in their output, corporate earnings and consequently in their capital markets, see, among others, Phylaktis and Ravazzolo (2002,2004). Secondly, the important role, especially for the markets included in this study, of country funds and alternative financial instruments that provide easier access for domestic and international investors to national markets and thus increase their financial links with world markets. Finally, markets deregulation and liberalization, technological developments in communications and trading systems, innovation in financial products and services and the increase in the international activities of multinational corporations can further induce relationshpis among national capital markets.

## 5 CONCLUSION

In this paper, we test a partially segmented ICAPM using an asymmetric extension of the multivariate GARCH process of De Santis and Gerard (1997,1998) for two developed countries (the US and UK), two emerging countries (Hong Kong and Singapore) and World market over the period February 1970- December 2003. This fully parametric empirical methodology, with sign and size asymmetric effects, allows to the prices of domestic and world market risks, betas and correlations to vary asymmetrically through time. The evidence shows that this asymmetric process provides a significantly better fit of the data than a standard symmetric process. Then, we test different pricing restrictions of the model over the whole period. The evidence supports the financial integration hypothesis and indicates that domestic risk is not a priced factor. Taking into account the fact that financial integration is an ongoing process, we re-estimate the model over two sub-periods 1970-1987 and 1987-2003. For the first sub-period, we find weak support for the hypothesis that the Hong Kong stock market is partially segmented and then investors in Hong Kong face both common and country-specific risks and price them both. For the other analysed markets, our findings support strongly the full integration hypothesis. Concerning the sub-period 1987-2003, the results of this paper show that thre is no evidence of financial segmentation and that the stock markets analysed are all subject to worlwide influences.

## REFERENCES

- Adler, M. and B. Dumas, 1983. International portfolio selection and corporation finance: a synthesis. *Journal of Finance*, 38, 925-984.
- Aggarwal, R, B. Lucey and C. Muckley, 2003. Dynamics of equity market integration in Europe: evidence of changes over time and with events. *Pronencia. International Symposium on International Equity Market Integration.* Trinity College. Dublin, juin 2003.
- Arshanapalli, B and J. Doukas, 1993. International stock market linkages: evidence from the pre- and the post-october 1987 period. *Journal of Banking and Finance*, 17, 193-208.
- Barari, M, 2003. Integration using time-varying integration score: the case of Latin America, *International Symposium on International Equity Market Integration*, Trinity College, Dublin, June 2003.
- Barr, D. and R. Priestley, 2004. Expected returns, risk and the integration of international bond markets. *Journal of International Money and Finance*, 23, 71-97.
- Bekaert, G, 1995. Market integration and investments barriers in emerging equity markets. *World Bank Economic Review*, 9, 75-107.
- Bekaert, G. and C. Harvey, 1995. Time varying world market integration. *Journal of Finance*, 50 (2), 403-444.
- Bekaert, G. and C. Havey, 2000. Foreign speculation and emerging equity markets. *Journal of Finance*, 55, 565-613.
- Bilson, C., V. Hooper, and M. Jaugietis, 2000. The impact of liberalisation and regionalism upon cpital markets in emerging Asian economies. *International Finance Review*, 1, 219-255.
- Bollerslev, T. and J. M. Wooldrige, 1992. Quasi-maximum likelihood estimation and inference in dynamic models with time-varying covariances. *Econometric Review*, 11, 143-172.
- Carrieri, F, V. Errunza and K. Hogan, 2002. Characterizing world market integration through time. *Working Paper,* McGill University.
- DE Santis, G, B. Gerard and P. Hillion, 2003. The relevance of currency risk in the EMU. *Journal of Economics and Business*, 55, 427-462.
- De Santis, G. and B. Gerard, 1997. International asset pricing and portfolio diversification with time varying risk. *Journal of Finance, 52*, 1881-1912.
- De Santis, G. and B. Gerard, 1998. How big is the premium for currency risk. *Journal of Financial Economics, 49, 375-412. .*
- Dumas, B. and B. Solnik, 1995. The world price of foreign exchange risk? Journal of Finance, 50, 445-479.
- Ferson, W. and C. Harvey, 1993. The risk and predictability of international equity returns. Review of Financial Studies, 6, 527-566.
- Gerard, B., K. Thanyalakpark and J. Batten, 2003. Are the East Asian markets integrated? Evidence from the ICAPM", *Journal of Economics and Business*, 55.

- Glosten, L., R. Jagannathan and D. Runkle, 1993. Relationship between the expected value and the volatility of national excess return on stocks. *Journal of Finance*, 48, 1779-1801.
- Hardouvelis, G, Malliaropoulos AND D. Priestley, 2002. EMU and stock market integration. *Working Paper*.
- Harvey, C, 1991. The world price of covariance risk. *Journal of Finance*, 46(1), 111-157.
- Hodrick, R. and E. Prescott, 1986. Post-War US business cycles: a descriptive empirical investigation. *Mark Watson,* Federal Reserve Bank of Chicago.
- Hunter, D. 2004. The evolution of stock market integration on the post-liberalization period- A look t Latin America. *Journal of International Money and Finance*. In press.
- Jeon, B. and G. Von Furstenberg, 1990. Growing international co-movement in stock price indexes. *Quarterly Review of Economics and Business*, 30, 15-31.
- Karolyi, A. and R. Stulz, 2002. Are financial assets priced locally or globally? *Working Paper,* Ohio State University.
- Kroner K. and V. Ng, 1998. Modelling asymmetric comovements of asset returns. *Review of Financial Studies*, 11, 817-844.
- Lau, S, T. and T, H. Mcinish, 1993. Comovements of international equity returns:a comparison of the pre and post-October 1987, periods. *Global Finance Journal*, 4, 1-19.
- Lim, K., H. Lee and K. Liew, 2003. International diversification benefits in Asian stock markets: a revisit. *Mimeo*, Lebuan School of International Business and Finance, Universiti Malasya Sabbah y Faculty of Economics and Management, University Putera Malaysia.
- Liu, Y, A. and M, S. Pan, 1997. Mean and volatility spillover effets in the US and Pacific-Basin stock markets", *Multinational Finance Journal*, 1, 47-62.
- Masih, A. and R. Masih, 1997. A Comparative analysis of the propagation of stock market fluctuations in alternative models of dynamic causal linkages. *Applied Financial Economics*, 7, 59-74.
- Masih, A. and R. Masih, 1999.Are Asian stock market fluctuations due mainly to intra-regional contagion effets? Evidence based on Asian emerging stok markets. *Pacific-Basin Finance Journal*, 7, 252-82.
- Masih, A. and R. Masih, 2001. Long and short term dynamic causal transmission amongst international stock markets. *Journal of International Money and Finance*, 20, 563-587.
- Merton, R, 1980. On estimating the expected return on the market: an explonary investigation. *Journal of Financial Economics*, 8(4), 323-361.
- Phylaktis, K. and F. Ravazzolo, 2000. Stock prices and exchange rate dynamics. *Mimeo*, City University Business School.
- Phylaktis, K. and F. Ravazzolo, 2002. Measuring financial and economic integration with equity prices in emerging markets. *Journal of International, Money and Finance*, 21, 879-903.
- Phylaktis, K. and F. Ravazzolo, 2004. Stock market linkages in emerging markets: implication for international portfolio diversification. Forthcoming, *Journal of International Markets and Institutions*.
- Ratanapakon, O. and S. Sharma, 2002. Interrelationships among regional stock indices. *Review of Financial Economics*, 11, 91-108.
- Roca, E. and E. Selvanathan, 2001. Australian and the three little dragons: are their equity markets interdepedent? *Applied Economic Letters*, 8, 203-07.

**Table 1: Descriptive statistics of asset excess returns**

**Panel A: Summary Statistics**

|  | Singapore | U.K. | H. Kong | U.S. | World |
|---|---|---|---|---|---|
| **Mean (% per year)** | 8.97 | 7.59 | 13.73 | 5.45 | 4.93 |
| **Min (% per year)** | -41.87 | -22.06 | -43.97 | -21.76 | -17.50 |
| **(date)** | (Oct. 1987) | (Oct. 1987) | (Oct. 1987) | (Oct. 1987) | (Oct. 1987) |
| **Max (% per year)** | 52.72 | 55.87 | 87.39 | 17.18 | 14.17 |
| **(date)** | (Jan. 1975) | (Jan. 1975) | (Fab. 1973) | (Oct. 1974) | (Jan. 1975) |
| **Std. Dev. (% per year)** | 111.28 | 85.59 | 140.12 | 54.73 | 51.26 |
| **Skewness** | 0.51* | 1.34* | -0.33* | 0.29** | -0.39* |
| **Kurtosis**[(1)] | 5.38* | 11.48* | 2.33* | 1.68* | 1.22* |
| **J.B.** | 499.76* | 2312.06* | 98.38* | 53.18* | 34.35* |
| **Q(12)** | 14.56 | 15.83 | 22.03** | 9.19 | 13.52 |

**Panel B: Unconditional correlations of $r_{it}$**

|  | Singapore | U.K. | H. Kong | U.S. | World |
|---|---|---|---|---|---|
| **Singapore** | 1.00 | 0.49 | 0.54 | 0.48 | 0.56 |
| **U.K.** |  | 1.00 | 0.38 | 0.55 | 0.69 |
| **H. Kong** |  |  | 1.00 | 0.35 | 0.51 |
| **U.S.** |  |  |  | 1.00 | 0.85 |
| **World** |  |  |  |  | 1.00 |

**Panel C: Autocorrelation of $(r_{it})$**

| Lag | Singapore | U.K. | H. Kong | U.S. | World |
|---|---|---|---|---|---|
| 1 | 0.093 | 0.080 | 0.089 | 0.015 | 0.075 |
| 2 | 0.009 | -0.093 | -0.011 | -0.031 | -0.051 |
| 3 | -0.078 | 0.062 | -0.023 | 0.018 | 0.044 |
| 4 | 0.045 | 0.009 | -0.092 | -0.041 | -0.027 |
| 5 | 0.019 | -0.109** | -0.071 | 0.098 | 0.077 |
| 6 | -0.062 | -0.052 | -0.033 | -0.042 | -0.032 |

**Panel D: Autocorrelation of $(r_{it})^2$**

| Lag | Singapore | U.K. | H. Kong | U.S. | World |
|---|---|---|---|---|---|
| 1 | 0.183* | 0.165* | 0.048 | 0.105** | 0.066 |
| 2 | 0.047 | 0.098 | 0.075 | 0.073 | 0.072 |
| 3 | 0.032 | 0.088 | 0.092 | 0.125** | 0.005 |
| 4 | 0.083 | 0.022 | 0.115** | 0.009 | 0.020 |
| 5 | 0.076 | 0.115** | 0.073 | 0.011 | 0.053 |
| 6 | 0.077 | 0.011 | 0.152** | 0.035 | 0.055 |

**Panel E: Cross-correlations of $(r_{it})^2$ - World and Country j**

| Lag | Singapore | U.K. | H. Kong | U.S. |
|---|---|---|---|---|
| -6 | -0.005 | -0.019 | 0.001 | -0.033 |
| -5 | -0.013 | -0.071 | 0.024 | 0.909 |
| -4 | 0.009 | 0.001 | 0.025 | -0.036 |
| -3 | 0.052 | 0.011 | 0.056 | 0.012 |
| -2 | -0.0131 | -0.053 | -0.001 | -0.055 |
| -1 | 0.055 | 0.035 | 0.063 | 0.017 |
| 0 | 0.562* | 0.695* | 0.409* | 0.859* |
| 1 | 0.064 | 0.063 | 0.031 | 0.045 |
| 2 | -0.032 | -0.042 | -0.019 | -0.007 |
| 3 | -0.029 | 0.023 | -0.063 | 0.079 |
| 4 | -0.023 | 0.016 | -0.036 | -0.056 |
| 5 | -0.017 | 0.070 | 0.009 | 0.009 |
| 6 | -0.107** | -0.053 | -0.025 | -0.053 |

*, ** Denote statistical significance at the 1% and 5%, (1) centred on 3.

**Table 2 : Asymmetric versus symmetric model**

$$\widetilde{R}_t - R_{ft}\tau = \delta_{t-1}h_{Nt} + \delta_{d,t-1}*q_t + \widetilde{\varepsilon}_t \qquad \widetilde{\varepsilon}_t/\Omega_{t-1} \sim N(0,H_t)$$

$$\delta_{t-1} = \exp(\kappa'_W Z_{t-1}) \,;\, \delta_{di,t-1} = \exp(\kappa'_i Z^i_{t-1})$$

**Symmetric model**
$$H_t = C'C + aa'*\varepsilon_{t-1}\varepsilon'_{t-1} + bb'*H_{t-1}$$

**Asymmetric model**
$$H_t = C'C + aa'*\varepsilon_{t-1}\varepsilon'_{t-1} + bb'*H_{t-1} + ss'*\xi_{t-1}\xi'_{t-1} + zz'*\eta_{t-1}\eta'_{t-1}$$

$$\xi_{it} = \varepsilon_{it} I_{\xi_{it}} \text{ where } I_{\xi_{it}} = 1 \text{ if } \varepsilon_{it} < 0 \text{ otherwise } I_{\xi_{it}} = 0,$$

$$\eta_{it} = \varepsilon_{it} I_{\eta_{it}} \text{ where } I_{\eta_{it}} = 1 \text{ if } |\varepsilon_{it}| > \sqrt{h_{iit}} \text{ otherwise } 0,$$

**Panel A: Likelihood ratio test**

| Null hypothesis | $\chi^2$ | df | p-value |
|---|---|---|---|
| $H_0$: $s = z = 0$ | 29.646 | 10 | 0.001 |

**Panel B : Information criterions**

| | Symmetric model | Asymmetric model |
|---|---|---|
| *AIC* | -11930.71 | -11963.03 |
| *SBC* | -11788.23 | -11815.59 |

**Panel C: Residual diagnostics**

| | Singapore | U.K. | H. Kong | U.S. | World |
|---|---|---|---|---|---|
| **Symmetric GARCH** | | | | | |
| *Mean($\times 100$)* | 0.27 | 0.06 | 0.29 | -0.02 | -0.11 |
| *Skewness* | 0.48* | 1.17* | -0.33** | -0.34* | -0.42* |
| *Kurtosis*[1] | 5.06* | 10.48* | 2.25* | 1.61* | 1.16* |
| *J.B.* | 441.59* | 1879.87* | 83.83* | 50.66* | 34.31* |
| *Q(12)* | 14.68 | 15.53 | 20.64 | 10.98 | 12.45 |
| **Asymmetric GARCH** | | | | | |
| *Mean($\times 100$)* | 0.12 | 0.003 | 0.09 | -0.02 | 0.05 |
| *Skewness* | 0.46* | 1.15* | -0.24** | -0.33* | -0.42* |
| *Kurtosis*[1] | 4.92* | 10.36* | 2.17* | 1.73* | 1.15* |
| *J.B.* | 408.19* | 1850.63* | 82.77* | 55.91* | 33.89* |
| *Q(12)* | 14.66 | 14.09 | 17.38 | 9.13 | 12.43 |

*\*, \*\* Denote statistical significance at the 1% and 5% , (1) centred on 3.*

## Table 3 : Quasi-maximum likelihood estimates of the partially integrated conditional ICAPM

$$\tilde{R}_t - R_{ft}\tau = \delta_{t-1}h_{Nt} + \delta_{d,t-1} * q_t + \tilde{\varepsilon}_t \qquad \tilde{\varepsilon}_t / \Omega_{t-1} \sim N(0, H_t)$$

$$\delta_{t-1} = \exp(\kappa'_W Z_{t-1}) \; ; \; \delta_{di,t-1} = \exp(\kappa'_i Z^i_{t-1})$$

$$H_t = C'C + aa' * \varepsilon_{t-1}\varepsilon'_{t-1} + bb' * H_{t-1} + ss' * \xi_{t-1}\xi'_{t-1} + zz' * \eta_{t-1}\eta'_{t-1}$$

$$\xi_{it} = \varepsilon_{it} I_{\xi_{it}} \text{ where } I_{\xi_{it}} = 1 \text{ if } \varepsilon_{it} < 0 \text{ otherwise } I_{\xi_{it}} = 0,$$

$$\eta_{it} = \varepsilon_{it} I_{\eta_{it}} \text{ where } I_{\eta_{it}} = 1 \text{ if } |\varepsilon_{it}| > \sqrt{h_{iit}} \text{ otherwise 0,}$$

**A: parameter estimates-mean equations**
*(a) Price of world market risk*

|  | Const. | WDY | DUSTP | USDP | DWIR |
|---|---|---|---|---|---|
| **Price of market risk** | 0.448* | 0.832* | -0.484** | 0.787* | -0.471 |
|  | (0.063) | (0.022) | (0.215) | (0.259) | (0.416) |

*(b) Price of domestic risk*

|  | Const. | LDY | DLIR | DIP |
|---|---|---|---|---|
| **Singa. Domestic Price** | -0.513 | 1.263 | -0.098 | 1.569 |
|  | (1.456) | (2.063) | (0.369) | (2.366) |
| **British Domestic Price** | 0.987 | 3.025 | -0.535 | -1.569 |
|  | (1.012) | (4.414) | (1.632) | (2.235) |
| **Hong K. Domestic Price** | 0.839 | 1.236 | -1.023 | -1.525 |
|  | (1.569) | (2.855) | (2.303) | (3.253) |
| **American Domestic Price** | 0.639 | 1.3302 | -1.003 | 0.236 |
|  | (0.968) | (5.165) | (1.588) | (2.634) |

**Panel B: parameter estimates-Multivariate GARCH process**

|  | Singapore | U.K. | Hong Kong | U.S. | World |
|---|---|---|---|---|---|
| **a** | 0.314* | 0.206* | 0.278* | 0.225* | 0.286* |
|  | (0.034) | (0.029) | (0.022) | (0.053) | (0.056) |
| **b** | 0.517* | 0.753* | 0.717* | 0.706* | 0.705* |
|  | (0.332) | (0.053) | (0.050) | (0.083) | (0.037) |
| **s** | 0.112** | 0.009 | 0.009* | 0.015** | -0.005 |
|  | (0.068) | (0.011) | (0.003) | (0.008) | (0.012) |
| **z** | 0.051 | -0.026** | -0.005 | -0.017** | -0.027** |
|  | (0.062) | (0.013) | (0.009) | (0.007) | (0.015) |

**Panel C: Specification tests**

| Null hypothesis | $\chi^2$ | df | p-value |
|---|---|---|---|
| *Is the price of world market risk constant?* <br> $H_0: \delta_{m,j}=0 \quad \forall j>1$ | 27.35 | 4 | 0.000 |
| *Is the price of American domestic risk equal to zero?* <br> $H_0: \delta_{dUs,j}=0$ | 0.90 | 4 | 0.924 |
| *Is the price of Singa domestic risk equal to zero?* <br> $H_0: \delta_{dS,j}=0$ | 1.27 | 4 | 0.866 |
| *Is the price of Hong K. domestic risk equal to zero?* <br> $H_0: \delta_{dHK,j}=0$ | 5.04 | 4 | 0.283 |
| *Is the price of British domestic risk equal to zero?* <br> $H_0: \delta_{dUK,j}=0$ | 0.99 | 4 | 0.910 |
| *Are the prices of domestic risk jointly equal to zero?* <br> $H_0: \delta_{d,j}=0 \quad \forall j,k$ | 4.03 | 16 | 0.999 |
| *Are the s coefficients jointly equal to zero?* <br> $H_0: s_i=0 \quad \forall i$ | 19.63 | 5 | 0.001 |
| *Are the z coefficients jointly equal to zero?* <br> $H_0: z_i=0 \quad \forall i$ | 14.02 | 5 | 0.007 |

*, ** Denote statistical significance at the 1% and 5% levels, QML standard errors are reported in parentheses, (a) equal to 0 for the normal distribution. Inorder to preserve space, estimates of the intercep matrix C is not reported.

**Tableau 4 : Robustness tests**

$$E(\tilde{R}_{it}/\Omega_{t-1}) - R_{ft} = \alpha_i + \delta_{t-1} Cov(\tilde{R}_{it}, \tilde{R}_{Wt}/\Omega_{t-1}) + \delta_{di,t-1} Var(Res_{it}/\Omega_{t-1}) + \phi_i' Z_{t-1}^i, \quad \forall i$$

$$\delta_{t-1} = \exp(\kappa_W' Z_{t-1}) \; ; \; \delta_{di,t-1} = \exp(\kappa_i' Z_{t-1}^i)$$

$$H_t = C'C + aa' * \varepsilon_{t-1}\varepsilon_{t-1}' + bb' * H_{t-1} + ss' * \xi_{t-1}\xi_{t-1}' + zz' * \eta_{t-1}\eta_{t-1}'$$

$$\xi_{it} = \varepsilon_{it} I_{\xi_{it}} \text{ where } I_{\xi_{it}} = 1 \text{ if } \varepsilon_{it} < 0 \text{ otherwise } I_{\xi_{it}} = 0,$$

$$\eta_{it} = \varepsilon_{it} I_{\eta_{it}} \text{ where } I_{\eta_{it}} = 1 \text{ if } |\varepsilon_{it}| > \sqrt{h_{iit}} \text{ otherwise } 0,$$

| Null hypothesis | $\chi^2$ | df | p-value |
|---|---|---|---|
| *Are country-specific constants all eaqual to zero?* <br> $H_0: \alpha_i = 0 \quad \forall i$ | 1.29 | 4 | 0.862 |
| *Are the local information variable coefficients jointly equal to zero?* <br> $H_0: \phi_i = 0 \quad \forall i$ | 3.26 | 12 | 0.514 |

**Table 5 : Sub-sample analysis**

$$\tilde{R}_t - R_{ft}\tau = \delta_{t-1} h_{Nt} + \delta_{d,t-1} * q_t + \tilde{\varepsilon}_t \qquad \tilde{\varepsilon}_t / \Omega_{t-1} \sim N(0, H_t)$$

$$\delta_{t-1} = \exp(\kappa_W' Z_{t-1}) \; ; \; \delta_{di,t-1} = \exp(\kappa_i' Z_{t-1}^i)$$

$$H_t = C'C + aa' * \varepsilon_{t-1}\varepsilon_{t-1}' + bb' * H_{t-1} + ss' * \xi_{t-1}\xi_{t-1}' + zz' * \eta_{t-1}\eta_{t-1}'$$

$$\xi_{it} = \varepsilon_{it} I_{\xi_{it}} \text{ where } I_{\xi_{it}} = 1 \text{ if } \varepsilon_{it} < 0 \text{ otherwise } I_{\xi_{it}} = 0,$$

$$\eta_{it} = \varepsilon_{it} I_{\eta_{it}} \text{ where } I_{\eta_{it}} = 1 \text{ if } |\varepsilon_{it}| > \sqrt{h_{iit}} \text{ otherwise } 0,$$

| | | 1970-1987 | | 1987-2003 | |
|---|---|---|---|---|---|
| Null hypothesis | df | $\chi^2$ | p-value | $\chi^2$ | p-value |
| *Is the price of world market risk constant?* <br> $H_0: \delta_{m,j}=0 \quad \forall j > 1$ | 4 | 37.56 | 0.000 | 25.13 | 0.000 |
| *Is the price of American domestic risk equal to zero?* <br> $H_0: \delta_{dUs,j}=0$ | 4 | 1.23 | 0.872 | 0.48 | 0.975 |
| *Is the price of Singa domestic risk equal to zero?* <br> $H_0: \delta_{dS,j}=0$ | 4 | 2.51 | 0.642 | 1.27 | 0.866 |
| *Is the price of Hong K. domestic risk equal to zero?* <br> $H_0: \delta_{dHK,j}=0$ | 4 | 9.66 | 0.046 | 3.33 | 0.502 |
| *Is the price of Hong K. domestic risk constant?* <br> $H_0: \delta_{dHK,j}=0 \quad \forall j > 1$ | 3 | 5.24 | 0.155 | - | - |
| *Is the price of British domestic risk equal to zero?* <br> $H_0: \delta_{dUK,j}=0$ | 4 | 3.873 | 0.423 | 0.94 | 0.918 |
| *Are the prices of domestic risk jointly equal to zero?* <br> $H_0: \delta_{d,j}=0 \quad \forall j,k$ | 16 | 15.16 | 0.512 | 3.26 | 0.999 |

**Figure1 : World price of risk**

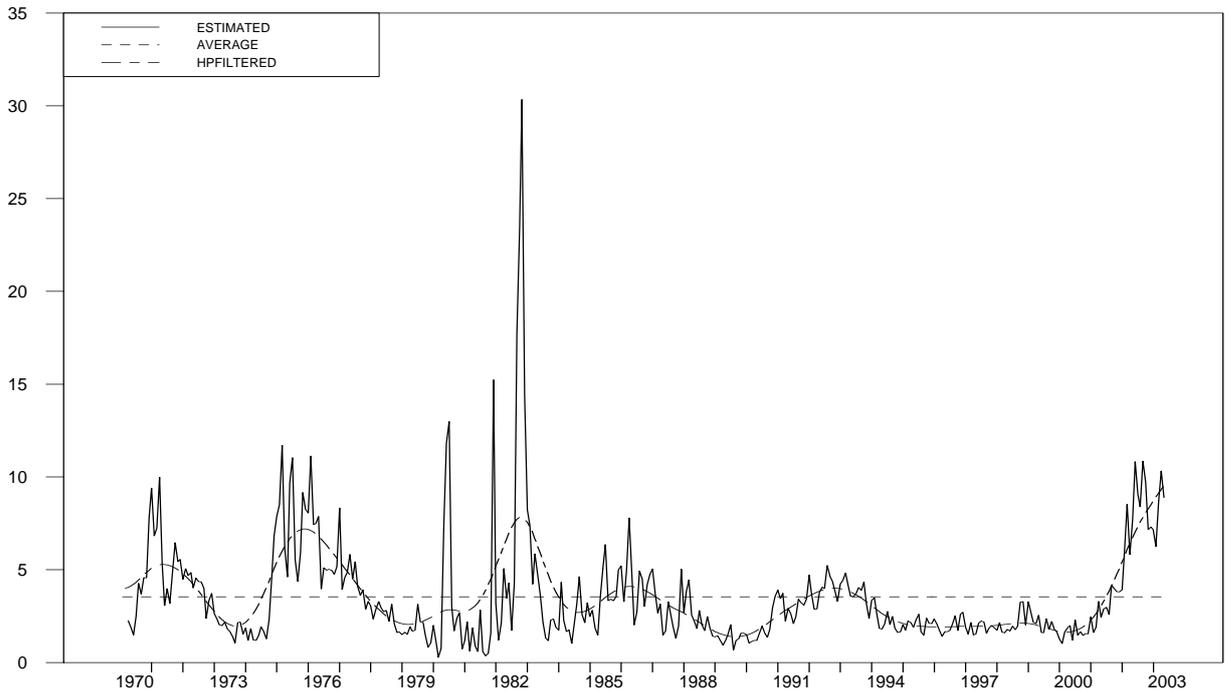

**Figure 2 : Conditional correlations with market portfolio**

*2-a Singapore*

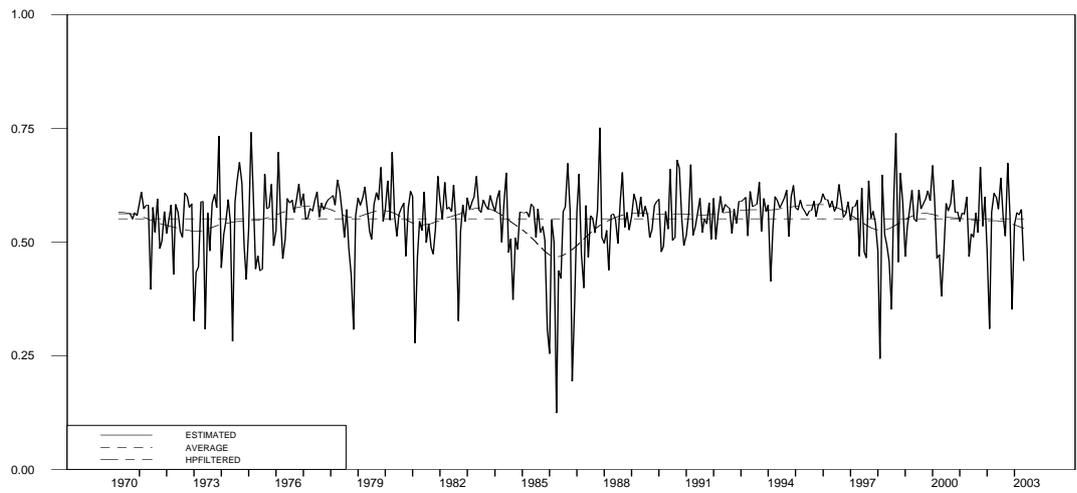

*2-b United Kingdom*

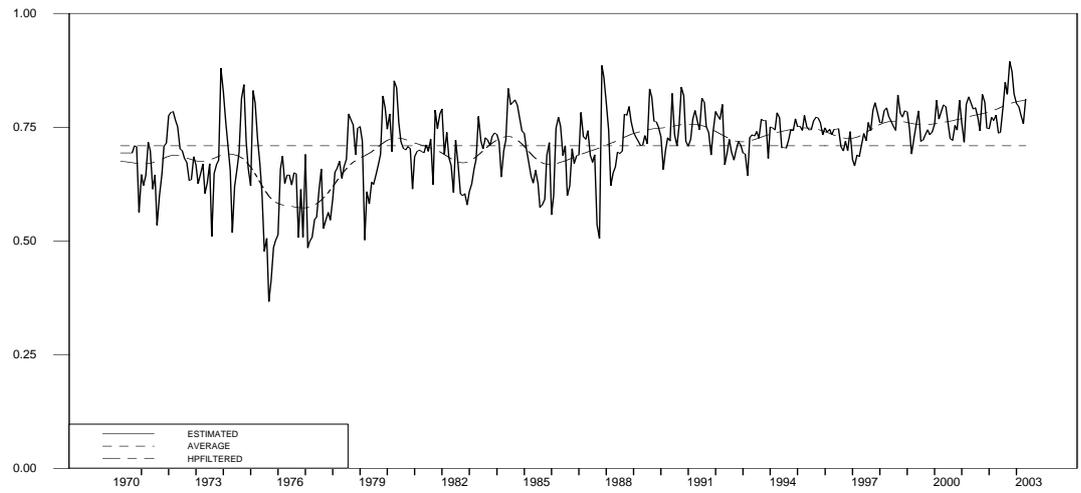

*2-c Hong Kong*

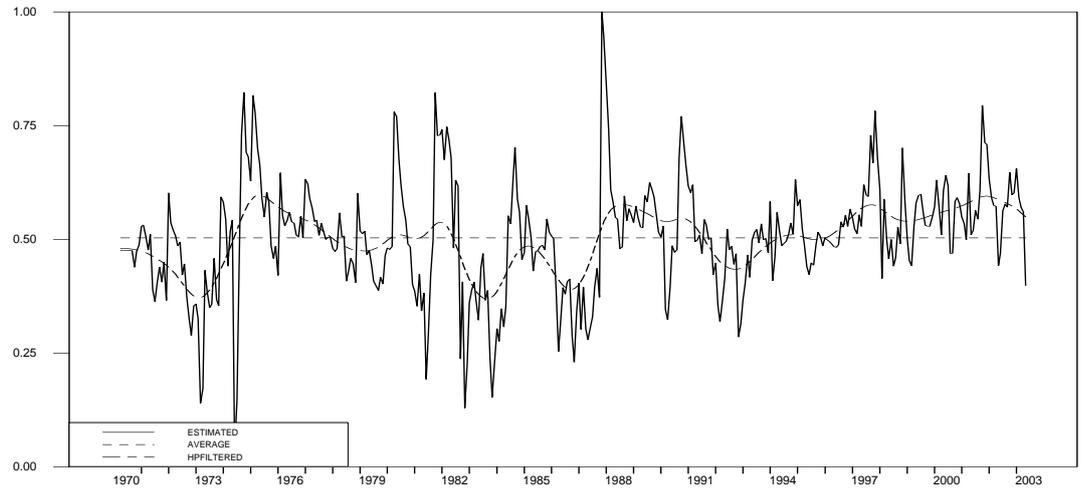

*2-d United States*

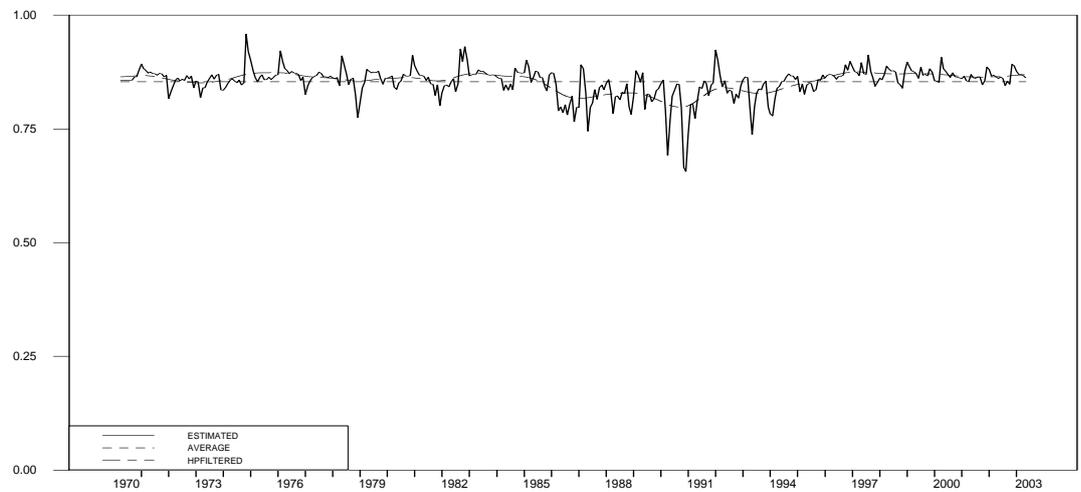

15